\documentclass[page-classic]{epl2}
\usepackage[final]{pdfpages}

\title{Dynamics of swollen fractal networks }

\author{ALVARO V. N. C. TEIXEIRA\footnote{Corresponding author at: Departamento de F\'{\i}sica - CCE, Universidade Federal de Vi\c cosa, 36570-900, Vi\c cosa/MG, Brazil. Tel.: +55 31 38992985. E-mail address: alvaro@ufv.br (A. V. N. C. Teixeira).} \and PEDRO LICINIO}
\shortauthor{Alvaro V. N. C. Teixeira \etal: DYNAMICS OF SWOLLEN FRACTAL NETWORKS}

\institute{                    
  Departamento de F\'{\i}sica - ICEx, Universidade Federal de Minas Gerais, C. P. 702, 30161-970, Belo Horizonte/MG, Brazil.\\
}

\abstract{
The dynamics of swollen fractal networks (Rouse model) has been studied through computer simulations. The fluctuation-relaxation theorem was used instead of the usual Langevin approach to Brownian dynamics. We measured the equivalent of the mean square displacement $\langle \vec r^{\,2} \rangle$ and the coefficient of self-diffusion $D$ of two-and three-dimensional Sierpinski networks and of the two-dimensional percolation network. The results showed an anomalous diffusion, i. e., a power law for $D$, decreasing with time with an exponent proportional to the spectral dimension of the network.}

\begin{document}

\maketitle

Free particles in solution present random trajectories that define their Brownian movement. This behavior is caused by the multiple collisions with the solvent molecules. The dynamics of this motion is well known and the diffusion law for these particles is given by 

\begin{equation}
\langle \vec r^{\,2} \rangle =2d_e Dt,
\label{eq1}
\end{equation} 

\noindent where $\langle \vec r^{\,2} \rangle$ is the mean square displacement of a particle, $D$ its diffusion coefficient and $d_e$ is the embedding space dimension. This asymptotic or Brownian regime is valid for times greater than the characteristic time of the collisions. For very short times (smaller than a few picoseconds) the particles move ballistically. 
Otherwise, when the particles interact or have their motion restricted by a distribution of barriers, their dynamics becomes altered at larger times so that the law of diffusion is not linear over the whole Brownian time range. The situations when the mean square displacements are ruled by a power law in time have been called anomalous diffusion. We can mention, as an example of anomalous diffusion, the random walk in a fractal substrate [1] with 

\begin{equation}
\langle \vec r^{\,2} \rangle \sim t^{d_s/d},
\label{eq2} 
\end{equation}

\noindent where $d_s$ is the spectral dimension and d is the fractal dimension of the substrate. Anomalous diffusion regimes, as well as logarithmic regimes, appear in random walks in bundled structures [2]. It can be also observed for particles whose movement is restricted by a static gel network [3]. 
It is interesting to note that this phenomenon is also seen in the self-diffusion of linear polymer units [4, 5] and we have recently shown that any ideal regular polymer network displays anomalous regimes [6, 7]. For these structures the diffusion coefficient 

\begin{equation}
D \equiv \frac{1}{2d_e} \frac{\partial \langle \vec r^{\,2} \rangle}{\partial t}
\label{eq3}
\end{equation}

\noindent is given by  
 
\begin{equation}
D = D_0 \left( \frac{t}{\tau} \right)^{-\alpha}
\label{eq4}
\end{equation}

\noindent where $D_0$ is the free diffusion coefficient of the units for a disconnected structure; $\tau$ is a characteristic time constant and $\alpha$ is the anomalous exponent which depends only upon the topological or graph dimension $d_t$ according to 

\begin{equation}
\alpha = \frac{d_t}{2}.
\label{eq5}
\end{equation}

This result (also obtained analytically by mode analysis [7]) shows that each dimension con­tributes with 1/2 in the anomalous exponent and the diffusion in each direction is independent of the others. 
As a result of the integration of (4), we have the diffusion regimes for structures with $d_t =1$ (linear chains - unbounded or nontrapped diffusion), $d_t = 2$ (two-dimensional networks - critical diffusion) and $d_t = 3$ (three-dimensional networks - bounded or trapped diffusion), respectively: 

\begin{equation}
\langle \vec r^{\,2} \rangle \sim t^{1/2},
\label{eq6}
\end{equation}
\begin{equation}
\langle \vec r^{\,2} \rangle \sim \ln (t/\tau),
\label{eq7}
\end{equation}
\begin{equation}
\langle \vec r^{\,2}_\infty \rangle - \langle \vec r^{\,2} \rangle \sim t^{-1/2},
\label{eq8}
\end{equation}

\noindent where $\langle \vec r^{\,2}_\infty \rangle$ is the asymptotic trapping mean square displacement.

\begin{figure}[h]
\centering
\includegraphics[scale=0.7,bb=50 30 557 260]{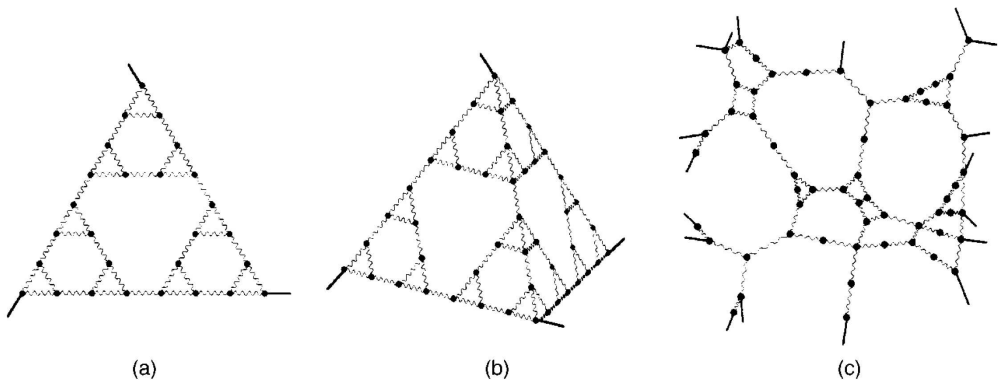}
\caption{Schematic diagram of: (a) two-and (b) three-dimensional Sierpinski networks and (c) relaxed two-dimensional percolation network. The continuous lines show the bounds relative to the periodic boundary conditions.}
\label{fig.1}
\end{figure}

In this letter we obtain generalized relations for the self-diffusion behavior of swollen fractal networks. To do this, we simulated two-and three-dimensional Sierpinski networks (figs. 1a and b) as well as a two-dimensional percolation network (fig. 1c), all of them following Rouse’s model, i. e., with units (network nodes) immersed in a viscous immobile solvent and bounded to their nearest neighbors by phantom springs. In this model the excluded volume and hydrodynamic interactions are neglected. The motivation of this work is understanding the dynamics of swollen networks as polymeric gels. Until now the studies concern only the diffusion of free particles in a static network, neglecting the network's own dynamics. 
The percolation network was constructed by cutting the bonds of an initially regular two-dimensional network with a probability of 50\%. Only the biggest cluster was considered and the smaller ones were excluded. In fig. 1c we represent the resulting relaxed percolation network. 
Just like our previous work [6], the network nodes are bound by springs with elastic constant k which represents linear polymers connecting the nodes in a gel, for example. This constant is given by 

\begin{equation}
k = \frac{3k_B T}{2\xi^2}.
\label{eq9}
\end{equation}

\noindent where $\xi$ is the Flory radius of a polymer filament and $k_B T$ has the usual meaning. 
In order to study network dynamics, we developed an alternative method based on the fluctuation-relaxation theorem. This theorem tells us that both the microscopical spontaneous fluctuations and the relaxation due to an external force follow the same rule. This means that, in our case, if we push one specific node of the network with force $\vec f$, we can obtain information about their Brownian movement. 
Thus, the simulations consist in pushing the $i$-th unit (or node) of the relaxed network with a constant force $\vec f$. The equation of motion for the $i$-th unit is 

\begin{equation}
\zeta \frac{\partial \vec r_i (t)}{\partial t} = -k \sum_j [\vec r_i (t) - \vec r_j (t)] + \vec f.
\label{eq10}
\end{equation}

The left term is the viscous drag on the $i$-th node which corresponds to a mean drag of all half-segments connecting to this node and the hydrodynamic drag $\zeta$ is

\begin{equation}
\zeta = \frac{k_B T}{D_0}.
\label{eq11}
\end{equation}

The first term to the right of (10) represents the spring interactions and the summation of $j$ is related to all immediate topological neighbors of the $i$-th particle. As in the method of Brownian dynamic simulations, or the Langevin equation, the inertial term was neglected. For the other units with no external force, eq. (10) is still used, but setting $\vec f =0$. 
It can be easily shown with the fluctuation-relaxation theorem that the displacement of the 
$i$-th node $[\vec r_i (t) - \vec r_i (0)]$ and the mean square displacement $\langle \vec r^{\,2}_i \rangle$, as well as the velocity of the $i$-th node $\vec v_i (t)$ and the diffusion coefficient $D_i (t)$, are intimately related [7] with 

\begin{equation}
[\vec r_i (t) - \vec r_i (0)] = \frac{\vec f}{2d_e k_B T} \langle \vec r^{\,2}_i \rangle
\label{eq12}
\end{equation}
 
\noindent and 

\begin{equation}
\vec v_i (t) = \frac{\vec f}{k_B T} D_i (t).
\label{eq13}
\end{equation}

The advantages of this method upon the traditional one is immediate: since there is no need for the calculation of mean values (the position in (12) is measured at once) and the generation of random numbers (to simulate the random forces), the computer memory and time needed are drastically reduced. On the other hand, if the network has an irregular structure (as the percolation network) then the mean square displacement has to be calculated by a mean of the displacements of each node with $\vec f$ applied to it. 
As a result of the simulations, we see that the mean square displacement per degree of freedom $\langle \vec r^{\,2} \rangle/2d_e$ calculated by (12) shows a normal behavior ($\sim t$) at short and long times (fig. 2 - in our simulations we use time units of $\xi^2/D_0$). The intermediate regime is anomalous (power law) and can be better observed in the normalized diffusion coefficient $D/D_0$ (fig. 3) calculated by (13). The constant diffusion limits are: free diffusion for short times ($D = D_0$) and center-of-mass diffusion for long times ($D = D_0/N$). 

\begin{center}
\begin{figure}[t]
\includegraphics[scale=0.75,bb=50 40 557 260]{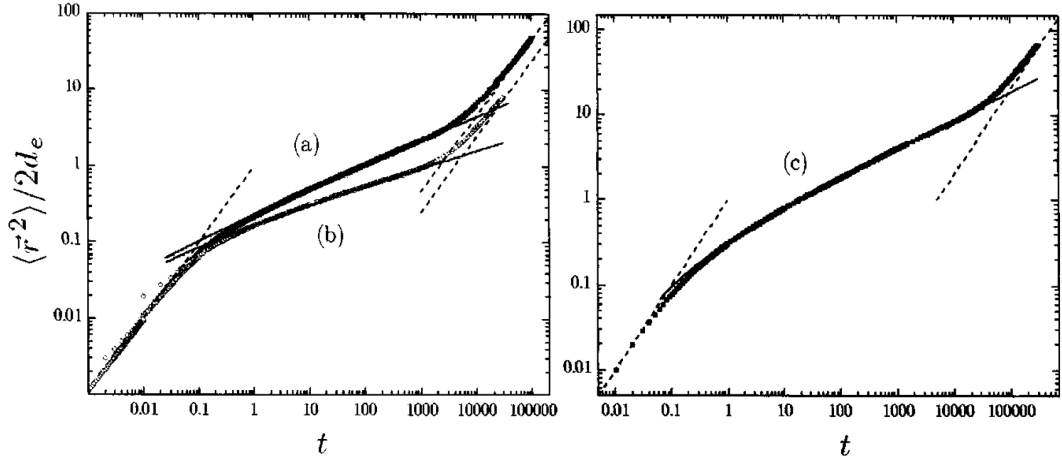}
\caption{Mean square displacement per degree of freedom for: (a) two-dimensional Sierpinski network ($N = 2187$ particles); (b) three-dimensional ($N = 4096$); (c) two-dimensional percolation network ($N = 4984$). The dashed line is the free diffusion regime ($\langle \vec r^{\,2} \rangle /2d_e = t$) and the diffusion of center of mass ($\langle \vec r^{\,2} \rangle/2d_e = t/N$). The continuous line is the fit by (16).}
\label{fig.2}
\end{figure}
\end{center}

\begin{center}
\begin{figure}[b]
\includegraphics[scale=0.75,bb=50 40 557 260]{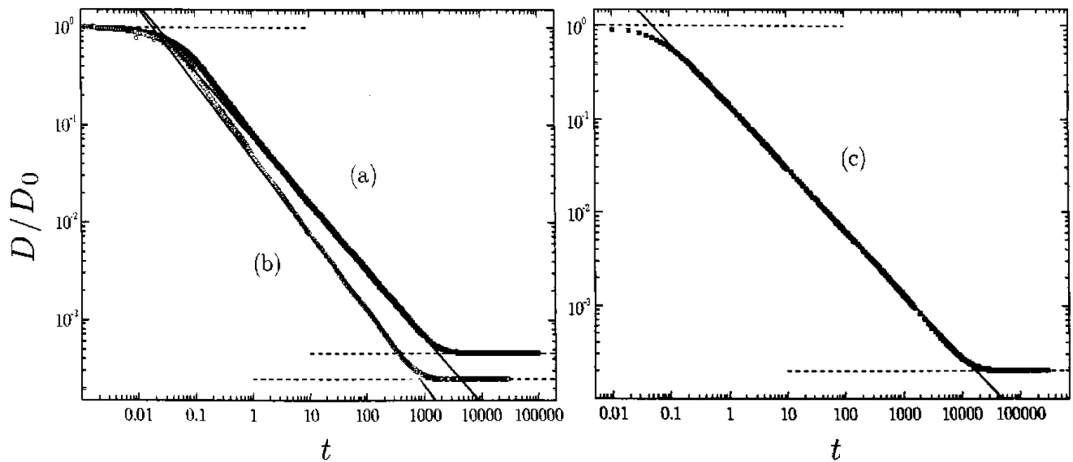}
\caption{Normalized diffusion coefficient for: (a) two-and (b) three-dimensional Sierpinski network and (c) two-dimensional percolation network. The continuous line is the fit by (4). The dashed line is the free diffusion regime and the diffusion of center of mass.}
\label{fig.3}
\end{figure}
\end{center}

\vspace{-2cm}
These results, including power law scaling, were similar to the ones found for regular networks - eq. (4) [6]. The anomalous exponent a was found to be related not to the graph dimension $d_t$, but to the spectral dimension $d_s$ by 

\begin{equation}
\alpha = \frac{d_s}{2}
\label{eq14}
\end{equation}

\noindent with deviations less than 3\% (the values of t and a obtained by fit of (4) are shown in table I). Since $d_s = d_t$ for regular networks, eq. (5) is a special case of (14). This general result indicates that the spectral dimension, originally used to describe the spectral density [8] of material with fractal structure, is also related with the self-diffusion of Rouse fractal networks. Unlike the problem of random walks in statical fractal substrates, eq. (2), relation (14) is yet to be proven. 

\begin{table}
\caption{Parameters resulting from fit of (4) to $D$ values. Also shown the values of half spectral dimension.}
\label{tab.1}
\begin{center}
\begin{tabular}{clll}
\hline
&&&\\
Network  & Sierpinski-2D & Sierpinski-3D & Percolation-2D \\
&&&\\
\hline
&&&\\
$\tau$   & 0.02492(1)    & 0.0218(2)     & 0.05527(8)  \\
&&&\\
$\alpha$ &  0.6895(2)    & 0.7912(9)     & 0.6753(3)  \\
&&&\\
$d_s/2$  & 0.6826        & 0.7737        & 0.6667  \\
&&&\\
\hline
\end{tabular}
\end{center}
\end{table}

\newpage
For Sierpinski networks the spectral dimension is given by [1] 

\begin{equation}
d_s = \frac{2 \log (d_e + 1)}{\log (d_e + 3)}.
\label{eq15}
\end{equation}

\noindent For percolation networks the spectral dimension is given in terms of critical exponents [1] and is estimated as $\simeq 4/3$ for any $d_e \geq 2$ according to the Alexander-Orbach conjecture [9]. 
Integrating (4) we have the mean square displacement of the structures with de differing 
from the critical dimension 2:

\begin{equation}
\frac{\langle \vec r^{\,2} \rangle}{2d_e} = \frac{\Delta}{2d_e} + \frac{D_0 \tau}{1-\alpha} \left( \frac{t}{\tau} \right)^{1-\alpha}.
\label{eq16}
\end{equation}

The values of $\tau$, $\Delta$ and $\alpha$ obtained by fit of (16) are given in table II. Note that in this table the negative values of the parameter . do not have any physical meaning outside the fitting range, which corresponds to the intermediate or anomalous diffusion regime ($\tau \ll t \ll \tau N^{1/\alpha}$).

\begin{table}
\caption{Parameters resulting from fit of (16) to $\langle \vec r^{\,2} \rangle$ values. Also shown the values of half spectral dimension.}
\label{tab.2}
\begin{center}
\begin{tabular}{clll}
\hline
&&&\\
Network  & Sierpinski-2D & Sierpinski-3D & Percolation-2D \\
&&&\\
\hline
&&&\\
$\tau$   & 0.02510(1)  & 0.0219(1)  & 0.0520(2)  \\
&&&\\
$\Delta$ & -0.1395(4)  & -0.418(7)  & -0.450(7)  \\
&&&\\
$\alpha$ & 0.6904(3)   & 0.7922(6)  & 0.6697(3)  \\
&&&\\
$d_s/2$  & 0.6826      & 0.7737     & 0.6667  \\
&&&\\
\hline
\end{tabular}
\end{center}
\end{table}

The dynamic self-structure factor can be easily obtained since the probability of displacement of particles in a given time follows a Gaussian statistics even in the anomalous regime (not shown here). So $S_s(q,t) = \exp[-q^2\langle \vec r^{\,2}\rangle/2d_e]$ and, using (16) and (14), we have 

\begin{equation}
S_s(q,t) \sim \exp \left[ -(\Gamma t)^{1-\alpha} \right],
\label{eq17}
\end{equation}

\begin{equation}
\Gamma = \left[(1-\alpha) \tau^\alpha q^2 \right]^{\frac {1}{1-\alpha}}.
\label{eq}
\end{equation}

\begin{center}
\begin{figure}[t]
\includegraphics[scale=0.45,bb=-150 40 557 600]{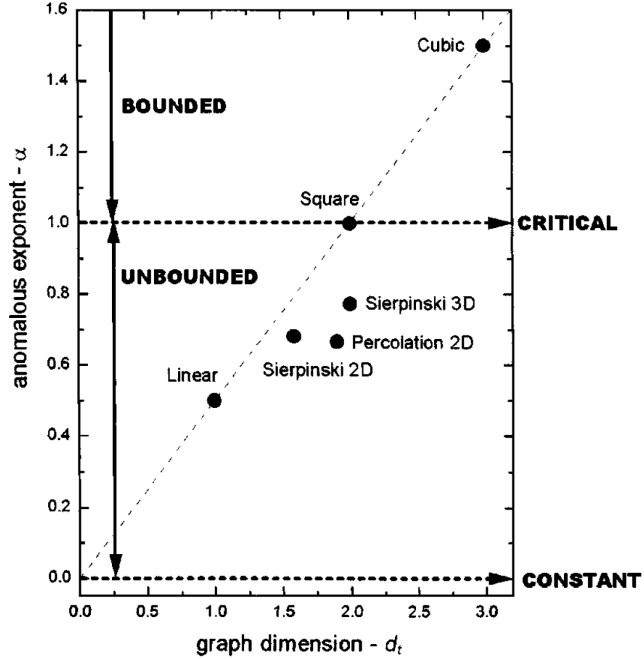}
\caption{Diagram of the self-diffusive regimes for Rouse networks. The fractal networks are in the unbounded region with the graph dimension equivalent to the Hausdorff dimension (1.59 and 2.00 for Sierpinski-2D and 3D, respectively, and 1.9 for percolation-2D). The regular networks (linear, square and cubic) lie in a line with inclination 1/2.}
\label{fig.4}
\end{figure}
\end{center}

With all these results we can plot a diagram of the Rouse network self-diffusive regimes (fig.4). The constant diffusion appears for $\alpha = 0$ for all $d_t$. The unbounded diffusion regime corresponds to region $0 < \alpha < 1$. The bounded regime corresponds to $\alpha > 1$ and the critical regime to $\alpha = 1$. 
The experimental counterpoint to these simulations could be given by the technique of dynamic light scattering or photon correlation applied to polymeric gels. One should like to extract information about the self-diffusion instead of the usual collective diffusion as we have done here. In that case, tracers should be attached to the network so that their scattering would overwhelm the gel background. Alternatively, inhomogeneous gels can contribute with a slow self-diffusion decay, easily detached from the fast collective decay as has been observed for polydisperse concentrated particle dispersion [10]. 
Perhaps the most simple way to obtain fractal gel is the percolation gel. For polymerization reactions in the presence of cross-linking agents it is expected that, when the solution is no longer fluid, or its viscosity diverges, the polymeric network reaches the percolation threshold or gel point [11]. In this condition the network is fractal with $d_s \simeq 0.67$. 
One severe limitation of our approach is the neglect of excluded volume and hydrodynamic interactions. We expect that these effects change results only in a quantitative way. Yet, the Rouse model could be approached with theta solvents and at high concentrations, where hydrodynamic interactions are well screened [12]. The effect of these interactions is to accel­erate the diffusion by increasing the anomalous exponent. It is probable that the anomalous exponent found in experiments is higher than the ones calculated here due to the presence of hydrodynamic interactions that cause an effect of increasing the diffusion. In this case the value of the exponent in (14) should be seen as an upper bound.

\acknowledgments
This work was supported by the Brazilian agency FAPEMIG.

\end{document}